\documentstyle[aps,preprint]{revtex}
\newcommand{\be}{\begin{equation}
\addtolength{\abovedisplayskip}{\extraspaces}
\addtolength{\belowdisplayskip}{\extraspaces}
\addtolength{\abovedisplayshortskip}{\extraspace}
\addtolength{\belowdisplayshortskip}{\extraspace}}
\newcommand{\ee}{\end{equation}}
\newcommand{\ba}{\begin{eqnarray}
\addtolength{\abovedisplayskip}{\extraspaces}
\addtolength{\belowdisplayskip}{\extraspaces}
\addtolength{\abovedisplayshortskip}{\extraspace}
\addtolength{\belowdisplayshortskip}{\extraspace}}
\newcommand{\ea}{\end{eqnarray}}
\newcommand{\newsection}[1]{
\vspace{7mm}
\pagebreak[3]
\addtocounter{section}{1}
\setcounter{subsection}{0}
\setcounter{footnote}{0}
\large {\bf\thesection. #1 \\}
\normalsize
\nopagebreak
\medskip
\nopagebreak
\hspace{3mm}}

\begin{document}
\addtolength{\baselineskip}{.7mm}
\thispagestyle{empty}
\begin{flushright}
IPM \# \\
1996
\end{flushright}
%
\begin{center}
{\Large{\bf Shape Phase Transition of Polyampholytes in \\
Two Dimensions}} \\[5mm]
{\bf M. R. Ejtehadi}\footnote{
\it e-mail: reza@netware2.ipm.ac.ir}
and
{\bf S. Rouhani}

{\it Department of Physics, Sharif University of Technology,}\\
{\it Tehran P. O. Box: 11365-5531, Iran.}\\
and \\
{\it Institute for Studies in Theoretical Physics and Mathematics}\\
{\it  Tehran P. O. Box: 19395-5746, Iran.}\\
{\parbox{14cm}{\hspace{5mm}
\begin{center}
{\bf Abstract}\\[5mm]
\end{center}
We have studied the transition in shape of
two dimensional polyampholytes using Monte Carlo
simulation. We observe that polymers with randomly
charged monomers get into a globular shape
at lower temperatures, provided that their total charge  is
below a critical value $Q_c$. Collapse into globular form happens
for all forms of force law, but the critical charge depends
on the form of the force law, inversely dependent on the
range of interaction. The value of the critical charge
is proportional to $N$, the size of the polymer, in the thermodynamic limit.
}}
\end{center}
\vfill
\newpage
\setcounter{section}{0}
\setcounter{equation}{0}
%
%
%
\newsection{Introduction}

The shape of polymers in equilibrium, determines
some of its physical, chemical and
biological properties,
thus predicting the shape for a given chemical composition of a
polymer is a challenging problem of polymer science. Some of the interesting
analytical approaches to this problem come from the theory
of spin glasses \cite{BY,MPV}, the replica trick \cite{SG1,SG2,PGT}, 
the random energy model \cite{BW,BOSW,GO}, 
and the mean field approximation \cite{GO,DS,SGS}.

Due to extremely complicated nature of forces involved,
computer simulations have been the more successful tool in this study \cite{B}.
Modeling polymers on basis of a self avoiding random walk (SAW)
using computer simulations can predict some elementary geometrical
quantities such as the end to end distance ($R_e$) or the radius
of gyration ($R_g$). For some large polymers these 
two quantities have the same scaling with respect to $N$, 
the number of monomers,
\begin{equation}
< R_{e}^{2} > \sim < R_{g}^{2} > \sim  N^{\nu}.
\label{PI}
\end{equation}
The universal scaling index $\nu$, can be obtained through
Monte Carlo simulations, which agrees well with Flory's
calculation of $\nu$ for different dimensions, for a
neutral SAW \cite{F,dG,ROMC} {(table 1)}.
Flory's calculations of $\nu$ are exact expect in $d=3$,
 which never the less is in good agreement with simulation (table 2).

Real polymers however have interactions among their monomers
and between the monomers and the solvent.
Thus they behave differently to the predictions of table 1.
However all polymers in a good solvent, independently of their
monomer interaction, behave like a SAW at high temperatures.
Thus observed values of $\nu$ at high temperatures
should correspond to those of table 1 \cite{dG}. In low
temperatures the effect of forces become important,
thus we expect different values of $\nu$.

In low temperatures the polymer should get into
configurations which minimize the potential at the same
time as satisfying the self avoiding constraint. For
short range attractive forces, such as the Van der Waals forces,
the polymer gets into a $d$ dimensional shape, where $d$ need
not be an integer. The index $\nu$ in this case is $1/d$.

This phase transition, which qualitatively happens at the
point where the thermal energy is comparable with the potential energy, is
 referred to as the $\theta$-point. This is the transition
of a polymer from a free globular shape to a collapsed
globular shape, determined by the minima of the potential.
In this paper we study the $\theta$ transition for two
dimensional polyampholytes, polymers with randomly charged
monomers. In a random distribution of charge the
total charge of the polymer need not be zero. If
the net total charge becomes proportional to the
size of polymer, we then expect a behavior similar
to polyelectrolytes, polymers with singly charged monomers.
Higgs and Joany \cite{HJ} have shown that polyampholytes in a good
solvent, which have total vanishing charge have a
smaller radius of gyration than uncharged polymers. They have
estimated the size of the polymer assuming screening of
the long range electrostatic force between monomers and
using the Debye-Huckel estimation of the free energy.
Polyampholytes with electrostatic interaction and
without the screening effect, were studied by Kantor 
{\it et al.} \cite{KLK,KK},
in three spatial dimensions.
Using scaling arguments and Monte Carlo simulations, they showed
that polyampholytes with zero net total charge, collapse into a
globular form below a critical temperature ($T_c$). They also
showed that a critical net charge $Q_c$ exists below which the polymer
collapses, just like uncharged polyampholytes, where as above
$Q_c$ the polymer gets into a rod shape just like a polyelectrolyte.
Of course for  $T>T_c$ both types, irrespective of their total charge,
behave like a neutral SAW.

We have addressed this problem for polyampholytes
in two dimensions. This question may be relevant when
considering polymers in a thin film solvent, adsorbed
onto a surface.
The restriction  of movement in two dimensions,
do give different results for a SAW in two dimensions.
But we do  not expect qualitatively different behavior
with regards to temperature or total charge. Indeed
the same qualitative behavior is observed, below a critical
temperature collapse into globular form happens.
There  also exists a critical total charge $Q_c$,
below which, the polymer collapses just like a zero
net charge polyampholytes, and above which
a rod shape is taken, similar to polyelectrolytes.

However the interesting question is the dependence of $Q_c$
on the form of the potential. For short range potentials $Q_c$
is proportional to $N$, the size of the polymer. We performed
our simulations for a few different forms of the potential,
and find agreement with an approximate expression for $Q_c$
\begin{equation}
Q_{c}^{2} \simeq (aN^2+bN) q_{_0}^{2}.
\end{equation}
Where $q_{_0}$ is the charge of the monomer, $N$ is the
size of the polymer, $a$ and $b$ are potential dependent
constants. We give an approximate derivation of eq. (2),
which is not dependent on the dimension of space,
so it should also hold in higher spatial dimensions.

%
%
%
\newsection{Simulations}

To simulate the motion of a polymer we have adapted Rose's
model \cite{K}. In this model a polymer, made up of $N$ monomers is
replaced with a SAW with $N-1$ steps on a square lattice.
Every vertex of the lattice through which the walk has passed,
represents a monomer. Since two
monomers can not occupy the same location in space,
a SAW is a representation of a configuration of the polymer.
To allow for effects of interaction, each site of the lattice
on the random walk is allowed to take a random
charge, which we restrict to be $\pm q_{_0}$ for simplicity.

To calculate quantities such as the radius of gyration
one needs to perform averages over ensembles of configurations.
We used the Metropolis \cite{B} method of importance sampling to
generate these ensembles. Let us assume $C$ is
the set of all possible configurations. We then construct a
set of maps $f_{\alpha}$ which transform the members of $C$
into each other:
\begin{eqnarray}
 f_{\alpha} & : & C \rightarrow C  \nonumber \\
 f_{\alpha}[s_i] & = & s_j \ \ \ \ \ \ \ \ s_i,s_j \in C.
\end{eqnarray}
At any given moment, a weight factor determines the
probability of the action of $f_{\alpha}$, using the
interaction hamiltonian. Thus starting with a given configuration,
the action of $f_\alpha$ takes us through the phase space
$C$ with the passage of time, generating the more probable
configurations first. An interesting question is {\sl what is the
most efficient set of $f_{\alpha}$, such that any two
members of $C$ can be transformed to each other with
a finite number of actions of $f_{\alpha}$}.
A very large set of $f_{\alpha}$ is not desirable
since the computer has to run through the set at each time
step, which slows down the  algorithm considerably \cite{K}.
On the other hand a very small set of $f_{\alpha}$ leaves
some configurations unreachable. Consider the
following set of moves:

{\it i} - Standard move

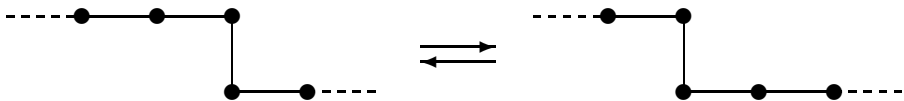
\setlength{\unitlength}{1mm}
\begin{picture}(120,14)(-42,0)
\multiput(0,12)(2,0){5}{\line(1,0){1}}
\put(10,12){\line(1,0){20}}
\put(30,12){\line(0,-1){10}}
\put(30,2){\line(1,0){10}}
\multiput(40,2)(2,0){5}{\line(1,0){1}}
\put(10,12){\circle*{2.00}}
\put(20,12){\circle*{2.00}}
\put(30,12){\circle*{2.00}}
\put(30,2){\circle*{2.00}}
\put(40,2){\circle*{2.00}}
\multiput(70,12)(2,0){5}{\line(1,0){1}}
\put(80,12){\line(1,0){10}}
\put(90,12){\line(0,-1){10}}
\put(90,2){\line(1,0){20}}
\multiput(110,2)(2,0){5}{\line(1,0){1}}
\put(80,12){\circle*{2.00}}
\put(90,12){\circle*{2.00}}
\put(90,2){\circle*{2.00}}
\put(100,2){\circle*{2.00}}
\put(110,2){\circle*{2.00}}
\thicklines
\put(55,8){\vector(1,0){10}}
\put(65,6){\vector(-1,0){10}}
\end{picture}

{\it ii} - End moves

\setlength{\unitlength}{1mm}
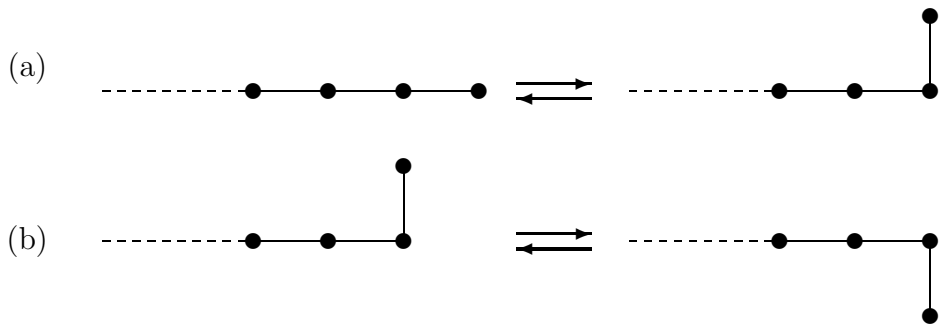
\begin{picture}(120,42)(-42,0)
\multiput(0,32)(2,0){10}{\line(1,0){1}}
\put(20,32){\line(1,0){30}}
\put(20,32){\circle*{2.00}}
\put(30,32){\circle*{2.00}}
\put(40,32){\circle*{2.00}}
\put(50,32){\circle*{2.00}}
\multiput(70,32)(2,0){10}{\line(1,0){1}}
\put(90,32){\line(1,0){20}}
\put(110,32){\line(0,1){10}}
\put(90,32){\circle*{2.00}}
\put(100,32){\circle*{2.00}}
\put(110,32){\circle*{2.00}}
\put(110,42){\circle*{2.00}}
\multiput(0,12)(2,0){10}{\line(1,0){1}}
\put(20,12){\line(1,0){20}}
\put(40,12){\line(0,1){10}}
\put(20,12){\circle*{2.00}}
\put(30,12){\circle*{2.00}}
\put(40,12){\circle*{2.00}}
\put(40,22){\circle*{2.00}}
\multiput(70,12)(2,0){10}{\line(1,0){1}}
\put(90,12){\line(1,0){20}}
\put(110,12){\line(0,-1){10}}
\put(90,12){\circle*{2.00}}
\put(100,12){\circle*{2.00}}
\put(110,12){\circle*{2.00}}
\put(110,2){\circle*{2.00}}
\thicklines
\put(55,33){\vector(1,0){10}}
\put(65,31){\vector(-1,0){10}}
\put(-10,35){\makebox(0,0){(a)}}
\put(55,13){\vector(1,0){10}}
\put(65,11){\vector(-1,0){10}}
\put(-10,12){\makebox(0,0){(b)}}
\end{picture}

\newpage
{\it iii} - Crank shaft move

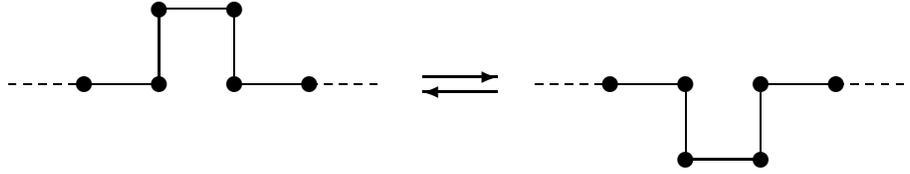
\setlength{\unitlength}{1mm}
\begin{picture}(120,24)(-42,0)
\multiput(0,12)(2,0){5}{\line(1,0){1}}
\put(10,12){\line(1,0){10}}
\put(20,12){\line(0,1){10}}
\put(20,22){\line(1,0){10}}
\put(30,22){\line(0,-1){10}}
\put(30,12){\line(1,0){10}}
\multiput(40,12)(2,0){5}{\line(1,0){1}}
\put(10,12){\circle*{2.00}}
\put(20,12){\circle*{2.00}}
\put(20,22){\circle*{2.00}}
\put(30,22){\circle*{2.00}}
\put(30,12){\circle*{2.00}}
\put(40,12){\circle*{2.00}}
\multiput(70,12)(2,0){5}{\line(1,0){1}}
\put(80,12){\line(1,0){10}}
\put(90,12){\line(0,-1){10}}
\put(90,2){\line(1,0){10}}
\put(100,2){\line(0,1){10}}
\put(100,12){\line(1,0){10}}
\multiput(110,12)(2,0){5}{\line(1,0){1}}
\put(80,12){\circle*{2.00}}
\put(90,12){\circle*{2.00}}
\put(90,2){\circle*{2.00}}
\put(100,2){\circle*{2.00}}
\put(100,12){\circle*{2.00}}
\put(110,12){\circle*{2.00}}
\thicklines
\put(55,13){\vector(1,0){10}}
\put(65,11){\vector(-1,0){10}}
\end{picture}

{\it iv} - End reflection

\setlength{\unitlength}{1mm}
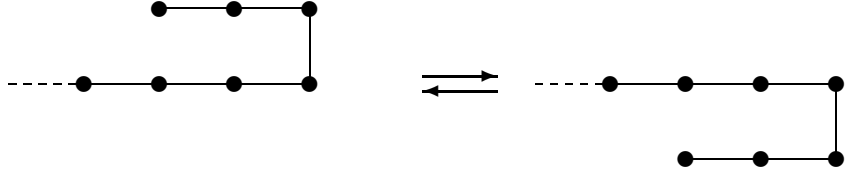
\begin{picture}(120,24)(-42,0)
\multiput(0,12)(2,0){5}{\line(1,0){1}}
\put(10,12){\line(1,0){30}}
\put(40,12){\line(0,1){10}}
\put(40,22){\line(-1,0){20}}
\put(10,12){\circle*{2.00}}
\put(20,12){\circle*{2.00}}
\put(30,12){\circle*{2.00}}
\put(40,12){\circle*{2.00}}
\put(40,22){\circle*{2.00}}
\put(30,22){\circle*{2.00}}
\put(20,22){\circle*{2.00}}
\multiput(70,12)(2,0){5}{\line(1,0){1}}
\put(80,12){\line(1,0){30}}
\put(110,12){\line(0,-1){10}}
\put(110,2){\line(-1,0){20}}
\put(80,12){\circle*{2.00}}
\put(90,12){\circle*{2.00}}
\put(100,12){\circle*{2.00}}
\put(110,12){\circle*{2.00}}
\put(110,2){\circle*{2.00}}
\put(100,2){\circle*{2.00}}
\put(90,2){\circle*{2.00}}
\thicklines
\put(55,13){\vector(1,0){10}}
\put(65,11){\vector(-1,0){10}}
\end{picture}

Without the moves ({\it ii}.b) and ({\it iv}), the following configuration:

\vspace{5mm}
\setlength{\unitlength}{1mm}
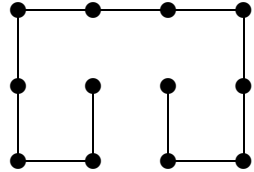
\begin{picture}(30,20)(-100,0)
\put(0,20){\line(1,0){30}}
\put(0,0){\line(0,1){20}}
\put(30,0){\line(0,1){20}}
\put(20,0){\line(0,1){10}}
\put(10,0){\line(0,1){10}}
\put(0,0){\line(1,0){10}}
\put(20,0){\line(1,0){10}}
\multiput(0,0)(10,0){4}{\circle*{2.00}}
\multiput(0,10)(10,0){4}{\circle*{2.00}}
\multiput(0,20)(10,0){4}{\circle*{2.00}}
\end{picture}
\newline
can not be reached.  However, the other moves may be sufficient 
for simulation of SAW, at high temperatures because 
the compact configurations such as the above  have very 
small entropy. But for our problem, 
where we seek compact configurations, at low temperature we need 
the full set of above moves.
Apparently this set of moves seems to be sufficient
for our problem, 
however we don't have a proof of this.

For a given polymer of size $N$, the algorithm running
time grows exponentially with $N$. For each configuration
$N$ attempts at changing the configuration is made.
The number of possible configurations grows like $N^2$
in two dimensions. Note that the algorithm has to check
the illegal moves as well. Thus the time to reach
equilibrium grows like $N^3$ . In practice we observed
that approximately $100N^3$ was necessary to reach equilibrium.
On the other hand, calculating the change in energy, going from
one configuration to another,
requires $N$ calculations due to the long range nature
of the force law.
Thus run time grows like $N^4$,
hence very long polymers are not easy to simulate.
We were thus restricted to a size $N\leq32$, with once
exception where we succeeded in getting result for $N=48$,
too.

Attributing a fixed charge $\pm q_{_0}$ to monomers at random,
leaves us with a polymer with a total  random charge,
between $+Nq_{_0}$, and $-Nq_{_0}$. We then  choose a
monomer at random and change its charge until the desired
total charge  is achieved. This gives the initial configuration to
start with. We then allow this configuration reach
equilibrium for a given temperature following the Metropolis method \cite{B},
applying the moves already discussed.
After reaching equilibrium the radius of gyration is calculated,
and this is then repeated $30$ times at large time intervals.
The whole procedure is then repeated with different charge configurations.
Averaging over charge configuration a final value for $<R_g>$ is
obtained. Repeating the above for different
value of $N$ gives a scaling relationship between
$<R_g>$ and $N$, giving a value for $\nu$.

%
%
%
\newsection{Results}

We simulated polyampholytes with size $N=\{32,16,8,4\}$ in
two spatial dimensions, using different force laws acting
 between monomers. We assumed an electrostatic interaction
of form $\log(r)$, $\frac{1}{r}$, $\frac{1}{r^2}$, and
a short range interaction $U(r)=0$ for $r \geq l$,
and constant otherwise. To begin with  we simulated polymers with
total charge zero, and observed that there exist a phase
transition as in three dimensions, where the polymer gets
into a globular shape at low temperatures.
This phase transition is observed for all types of potentials
used.

In figs. (1.a) and (1.b) the radius of gyration
is plotted {\it vs.} $\beta = (\frac {1}{T})$, for the two
potentials $\log(r)$ and  $\frac{1}{r}$. Where we observe a 
reduction in $R_g$ as $\beta$ increases. This reduction is better
seen for large polymers. We can observe
the changes in $\nu$ as a function of $\beta$ for these
two potentials in tables (3.a) and (3.b) and figs. (2.a)
and (2.b). 
We observe that this two types of polymers have
similar behavior and the value of $\nu$ for high temperatures is consistent
with the two dimensional neutral SAW. The interesting point is that
lower  temperatures, these polymers cannot collapse completely,
thus $\nu > 0.5$, which is the index for a two dimensional shape.
This may have been caused by considering an
incomplete set of moves, which can not access a collapsed shape.
However increasing the number of moves and repeating the simulations
and consistency of our results suggest otherwise. In two dimensions
holes can form which ore surrounded by like charges,
thus, movement to a more compact configuration becomes
energetically inaccessible. Probability of having a length $l$
of like charged monomers is proportional to a power of $l$ \cite{EK},
 thus the distribution  and size of these holes obeys a scaling law.
Therefore the existence of such holes affects the
scaling dimension of globular collapse in two dimensions and forces
$\nu$ above $0.5$. With purely attractive forces these
holes would not exist and this  obstruction is removed.

Repeating simulations for polymers with total nonzero
charge, showed that the phase transition observed by Kantor and Kardar
\cite{KK} in three dimensions, exists in two dimensions as well.
For a total charge greater than $Q_c$, rod shape is observed
where as for $Q < Q_c$, globular form, just like a zero net
charge is observed. The interesting point is the dependence of
$Q_c$ on the  potential and the size of the polymer.

In figs. (3a-3c) we observe the results for $<R_g>$, for a polymer
of  length $N=32$, using different potentials,
and in figs. (4a-4c) results for $N={48,32,16}$, using the potential
of the eq. (12). Although a
critical $Q_c$ dose exist for each potential, but the value of $Q_c$
depends inversely on the range of the potential.

%
%
%
\newsection{The Critical Charge}

The charge on monomers is a random variable of the problem but
without dynamics, thus it should be treated as a quenched 
variable when analyzing this systems. This is the root of 
similarity between this problem and spin glass systems.
The appropriate quenched average is to be effected as the 
free energy
\begin{equation}
F=\overline{\log Z}
\end{equation}
where bar means average over quenched variable, in our case
the different charge configuration. The calculation  
of quantities such as eq. (4) is riddled with difficulties,
one way is the replica method \cite{SG1,SG2}, but there is a lot of 
controversy associated with it. We were unable to perform a 
direct calculation
of eq. (4) for our system. Let us instead often a compromise 
quantity the weighted average potential energy:
\begin{equation}
<< U >>= \int \int D_r P(r) D_q P(q) \sum_{i,j} U(r_i,r_j) 
e^{-\beta U(r_i,r_j)}.
\end{equation}

In this expression $D_r P(r)$ integrates over all possible  self 
avoiding configurations including the appropriate distribution 
in space, $D_q P(q)$ like wise averages over 
all charge configurations, the factor $e^{-\beta U}$ allows 
the more probable energy configurations to happen more frequently 
in the averaging. The expression of eq. (5) lacks 
a normalizing factor which can be added later. 
The sum over $i$ and $j$, sums over all monomer 
pairs in interaction. Taking out the weight factor 
$e^{-\beta U}$  we end up with the expression used by Kantor 
and Kardar \cite{KK}. In this case the calculation of eq. (5) leads to:
\begin{equation}
<<U>> \sim (Q^2- Nq_{_0}^2)
\end{equation}
which indicates that the average potential changes sign 
at the point of total charge $Q$ equaling $q_{_0}\sqrt{N}$.
Thus a phase transition is expected at critical total 
charge of $q_{_0}\sqrt{N}$. This was observed by Kantor 
and Kardar but is inconsistent with our observations.

Not including the Boltzman  factor implies that two spatially
similar configuration, with different charge configurations 
are equally probable.  Where as clearly when opposite charges 
are near each other, a greater probability is achieved 
due to lower energy. Let us now keep the Boltzman factor 
in eq. (5). We can write eq. (5) as :
\begin{equation}
<< U >> = n_{+}\int D_r P(r) U(r) e^{-\beta U(r)} - n_{-}\int D_r P(r)
u(r) e^{-\beta U(r)}
\end{equation}
where $n_{+}$ and $n_{-}$ are the number of pairs of  
monomers with like and unlike charges respectively:
\begin{eqnarray}
n_{+} & = & \frac {N(N-2)q_{_0}^2 +Q^2}{4} \nonumber \\
n_{-} & = & \frac {N^2q_{_0}^2-Q_ý^2}{4}
\end{eqnarray}
which results in:
\begin{equation}
<<U>> \sim (Q^2-\frac {2N+(\lambda - 1)N^2}{\lambda + 1}q_{_0}^2)
\end{equation}
where
\begin{equation}
\lambda = \frac {\int D_rP(r)e^{\beta U(r)}}{\int D_rP(r)
e^{-\beta U(r)}}.
\end{equation}
The parameter $\lambda$, can only be determined if the 
shape of the potential is given, but $\lambda$ is positive 
and larger than $1$.
We now observe that the average potential energy changes 
sign at the critical charge:
\begin{equation}
Q_{c}^2 = [ (\frac {\lambda -1}{\lambda + 1})N^2 + \frac 
{2}{\lambda+1}N]q_{_0}^2.
\end{equation}
Thus the critical charge is proportional to $N$, rather than 
$\sqrt{N}$. This is in better agreement with our results. We 
can calculate $\lambda$ for a simple potential:
\begin{equation}
U(r)=\left\{ \begin{array}{ll}
		       V   & \ \ \ \ \  r \leq l \\
		0   & \ \ \ \ \  r>l \\
		\end{array}                     
     \right.
\end{equation}
We have $\lambda = e^{2\beta V}$. Comparison with fig. (4.b) 
indicates $16 \leq Q_c \leq 20$ for $N=32$, which is 
consistent with eq. (11). This result indicates that  in the 
thermodynamic limit where $N\rightarrow \infty$, critical 
charge scales with $N$. If the factor $\lambda$ calculates 
to be almost near $1$, then we get a different scaling 
behavior where $Q_c$ scales like $N^{\frac{1}{2}}$. This 
is of course true for high temperatures, i.e. $\lambda = 1$,  
but in this region, interactions don't play a role and 
all polymers behave like neutral self avoiding random walks.
     
%
%
%
\newsection{Discussion}

Simulating polyampholytes in two dimensions we observed the 
qualitative feature that a globular form is taken at low 
temperatures if the total charge of the polymer is less 
than a critical value $Q_c$. The scaling index $\nu$, of 
this globular form is above $1/2$, this is conjectured to 
be due to formation of areas surrounded by parts of 
the polymer made up of monomers 
of like charge. Where such an area is formed other parts of 
polymer are forbidden to enter it, due to the self avoiding 
nature of this system. Thus the globular form is more swollen 
than a two dimensional shape, hence $\nu > 1/2$. We were 
unable to predict the deviation from two dimensions by an 
analytic calculation, although this seems possible. Clearly 
a potential which is only attractive such as the Van der Waals 
potential would not give rise to this effect.

The other interesting observation concerns the scaling of $Q_c$ 
with size. Our analysis indicates that $Qþ_c$ scales with $N$. 
This becomes more stark when the forces are short range. 
It would also be experimentally observable if  the effects 
of  screening are made more intense, for example by changing the 
solvent \cite{inHJ,M}. It appears that this effect should exist 
independent of dimension, although our simulations were 
restricted to two dimensions. For a long range force the 
dependence of $Q_c$ on $N$ may become smaller, but should be important 
for large enough $N$. It must be said that our calculations 
were not based on a true quenched average. Since performing a true quenched 
average seems impossible, therefore some more convincing calculation  
is necessary before the question may be settled. 

{\bf Acknowledgments:} We are indebted to Mehran Kardar 
for valuable discussions and suggesting this problem to us.

%

%
%
\newcommand{\RMP}[1]{ Rev.\ Mod.\ Phys.\ {\bf #1}}
\newcommand{\PR}[1]{ Phys.\ Rev.\ {\bf #1}}
\small
%


\newpage
\begin{center}
\Large
Tables \\[10mm]
\normalsize
\end{center}

\begin{center}
\parbox{1.5in}{

\begin{center}
\begin{tabular}{|l|l|}  \hline
\ $d$ \       & \ Flory's $\nu$  \\ \hline
\ 1       & 1                  \\ \hline
\ 2       & $3/4$               \\ \hline
\ 3       & $3/5$              \\ \hline
\ $\geq4$   & $1/2$             \\ \hline
\end{tabular}   \\[10mm]
table 1. 
\end{center}
}
 \ \ \ \ 
\parbox{3in}{
\begin{center}
\begin{tabular}{|lr|}  \hline
 \ Monte Carlo Simulations       & for $d=3$ \\ \hline
 \ $0.59992\pm0.002$  &    \cite{G}   \\ \hline
 \ $0.5745\pm0.008\pm0.0056$& \cite{Fal}      \\ \hline
 \ $0.588\pm 0.001$   & \cite{LZ}   \\ \hline
\end{tabular}   \\[15mm]
table 2. 
\end{center}
}
\end{center}

{\bf table 1.} Flory's calculation of $\nu$ for different dimensions. 
All are exact except $d=3$.

{\bf table 2.} Monte Carlo simulation estimates of $\nu$ for $d=3$.

\begin{center} 
\parbox{2in}{
\begin{center}
$U(r)=\log(r)$   \\
\begin{tabular}{|l|l|l|}  \hline
\ \ $\beta$ & \ \ $\nu$ & \ \ $\Delta\nu$ \\ \hline
\ 0.02 \    & \ 0.761 \  & \ 0.009 \  \\ \hline
\ 0.08    & \ 0.753 & \ 0.009 \\ \hline
\ 0.32    & \ 0.726 & \ 0.010 \\ \hline
\ 1.28    & \ 0.644 & \ 0.013 \\ \hline
\ 5.12    & \ 0.614 & \ 0.016 \\ \hline
\ 20.48   & \ 0.60  & \ 0.012 \\ \hline
\end{tabular} \\[10mm]
table (3.a)           \\ 
\end{center} }
\ \ \ \ \ \ \ \
\parbox{2in}{
\begin{center}
$U(r)=\frac{1}{r}$ \\
\begin{tabular}{|l|l|l|}  \hline
\ \ $\beta$ & \ \ $\nu$ & \ \ $\Delta\nu$ \\ \hline
\ 0.02 \   & \ 0.762 \ & \ 0.008 \  \\ \hline
\ 0.08    & \ 0.761 & \ 0.009 \\ \hline
\ 0.32    & \ 0.744 & \ 0.010 \\ \hline
\ 1.28    & \ 0.672 & \ 0.017 \\ \hline
\ 5.12    & \ 0.590 & \ 0.032 \\ \hline
\ 20.48   & \ 0.61  & \ 0.036 \\ \hline
\end{tabular}           \\[10mm]
table (3.b).   \\

\end{center} }
\end{center}

{\bf Table 3.} The dependence of $\nu$ on $\beta$ for polymers with 
zero net charge for the potentials $\log(r)$ (3.a),
and $\frac{1}{r}$ (3.b).

\newpage
\begin{center}
\Large
Figure Captions \\[15mm]
\normalsize
\end{center}

{\bf Figure 1.}  

\hspace{3cm}
\parbox{12cm}{The radius of gyration of the polyampholyes
with zero net charge and different numbers of monomers {\it vs.} 
$\beta = 1/T$ for the potentials $\log(r)$ (1.a),
and $\frac{1}{r}$ (1.b).}

\vspace{5mm}

{\bf Figure 2.}  

\hspace{3cm}
\parbox{12cm}{ The dependence of $\nu$ on $\beta$ for polymers with 
zero net charge for the potentials $\log(r)$ (3.a),
and $\frac{1}{r}$ (3.b).}       

\vspace{5mm}

{\bf Figure 3.}  

\hspace{3cm}
\parbox{12cm}{The radius of gyration of the polyampholyes
with net charge $Q$ and $N=32$  {\it vs.} 
$\beta$ for the potentials $\log(r)$ (3.a),
$\frac{1}{r}$ (3.b), and $\frac{1}{r^2}$ (3.c).}

\vspace{5mm}

{\bf Figure 4.}  

\hspace{3cm}
\parbox{12cm}{The radius of gyration of the polyampholyes
with net charge $Q$ for the potential of eq. 12  
{\it vs.} $\beta$ for $N=16$ (4.a),
$N=32$ (4.b), and $N=48$ (4.c).}

\end{document}